\documentclass[modern]{aastex63}
\newcommand{\hi}{H\,{\sc i}}
\newcommand{\kms}{km\,s$^{-1}$}
\shorttitle{WAVES}
\shortauthors{Minchin R. et al.}
\begin{document}
\title{The Widefield Arecibo Virgo Extragalactic Survey: Early Results on Known Dark Sources}
\author[0000-0002-1261-6641]{Robert Minchin}\affiliation{SOFIA Science Center, Universities Space Research Association, MS 232-11, Moffett Field, CA 94035, USA}
\author[0000-0002-3782-1457]{Rhys Taylor}\affiliation{Astronomical Institute of the Czech Academy of Sciences, Bocni II 1401/1a, 141 00 Praha 4, Czech Republic}
\author[0000-0002-7898-5490]{Boris Deshev}\affiliation{Astronomical Institute of the Czech Academy of Sciences, Bocni II 1401/1a, 141 00 Praha 4, Czech Republic}
\correspondingauthor{Robert Minchin}\email{rminchin@sofia.usra.edu}
\begin{abstract}
The Widefield Arecibo Virgo Extragalactic Survey (WAVES) was an ongoing \hi\ survey of the Virgo Cluster with the Arecibo Observatory's 305-m William E. Gordon Telescope at the time of its structural failure. The full 20 square degrees of the southern field and 10 of the planned 35 square degrees of the northern field had been observed to full depth, adding to 25 square degrees observed to the same depth in the cluster by the Arecibo Galaxy Environment Survey (AGES). We here review what WAVES reveals about four optically-dark \hi\ structures that were previously discovered in the survey area, including two that are not seen despite being well above our detection limit
\end{abstract}

\section{Introduction}\label{intro}
WAVES was designed to build on the AGES observations of the Virgo Cluster \citep{2012MNRAS.423..787T,2013MNRAS.428..459T} and to take advantage of Arecibo's great brightness temperature sensitivity as a filled aperture telescope. Over a 10 \kms\ velocity width, AGES reached a $1\sigma$ column-density sensitivity of $1.5\times 10^{17} {\rm\ cm}^{-2}$ \citep{2016MNRAS.456..951K} while the shallower ALFALFA survey with Arecibo reached $5\times 10^{17} {\rm\ cm}^{-2}$ \citep{2008AA...487..161G}. The Arecibo surveys are  highly complementary to the VLA, where the VIVA survey of Virgo reached a relatively shallow $1-1.7\times 10^{19} {\rm\ cm}^{-2}$ \citep{2007ApJ...659L.115C} but had a 15\arcsec\ resolution compared to Arecibo's 3.5\arcmin\ beam. 

AGES discovered a large number of optically-dark \hi\ clouds, including some with anomalously high velocity widths that could not be easily explained as products of harassment \citep{2012MNRAS.423..787T,2017MNRAS.467.3648T}. The AGES data also demonstrated that deep surveys with good sensitivity to low column-density gas could detect previously undiscovered ram pressure stripped \hi\ tails: ten new tails were confidently detected, almost doubling the number of known tails in the cluster, along with sixteen possible detections \citep{2020AJ....159..218T}.

WAVES has the same integration time and observing method as AGES, and was intended to extend this column-density sensitivity to a large fraction of the Virgo cluster, either discovering new dark clouds and tails or putting strong upper limits on their population. The survey strategy was to complete small chunks of the survey to full sensitivity before moving on to the next section, thus 30 of a planned 55 square degrees was completed to full sensitivity before the Arecibo Telescope's collapse in 2020. A further 2.5 square degrees reached 2/3 of the planned integration time, i.e. 80\% of the planned sensitivity.

\section{Results}\label{results}

\begin{figure*}
\plotone{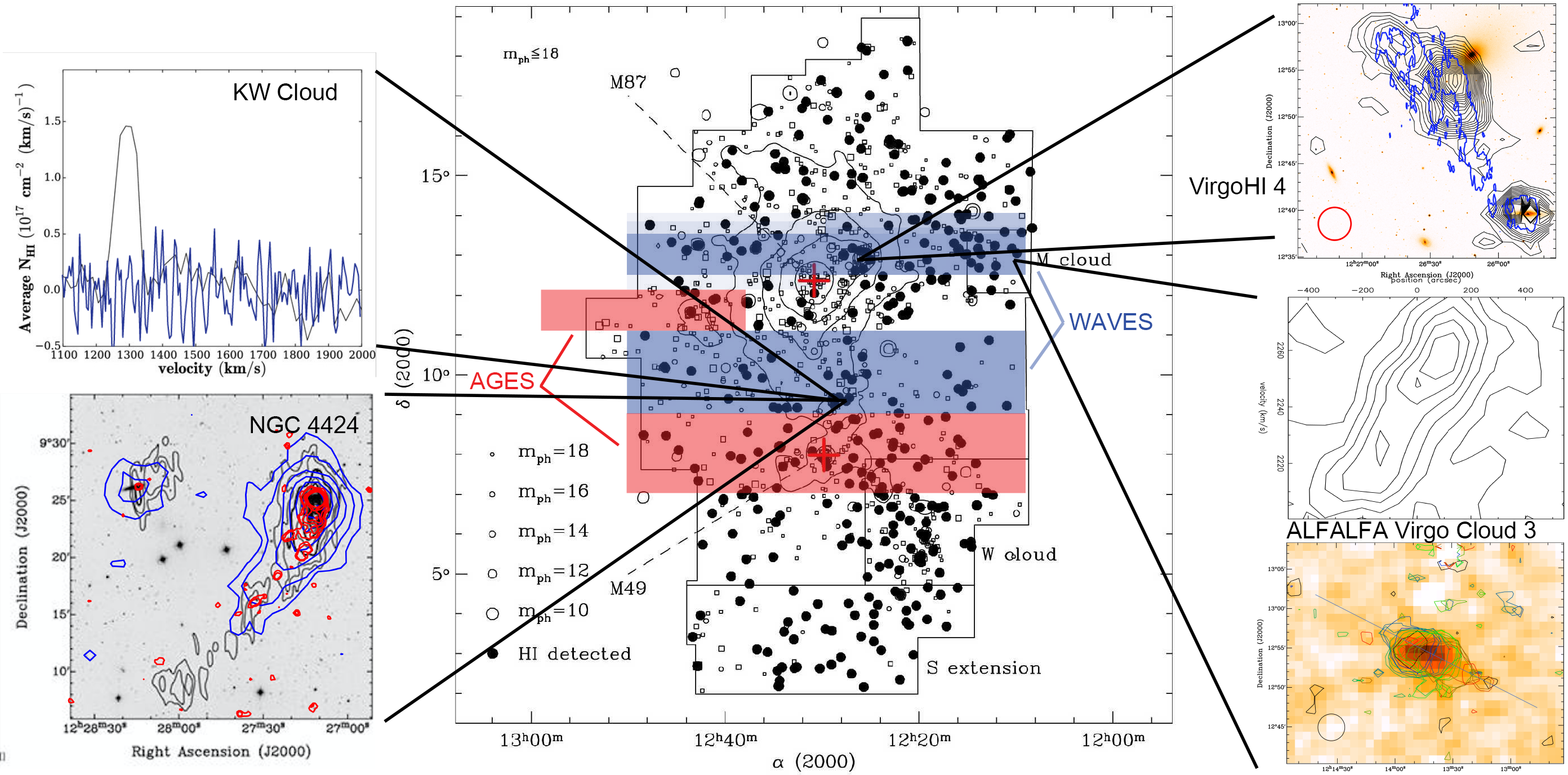}
\caption{WAVES (blue) and AGES (red) coverage of the Virgo cluster, with final integration times indicated by weight. Known dark \hi\ structures, clockwise from the top right: VirgoHI 4 \citep{2004MNRAS.349..922D} showing \hi\ contours from WAVES and the outer contour from the WSRT \citep[][blue]{2005AA...437L..19O} overlaid on the DSS. ALFALFA Virgo Cloud 3 \citep{2007ApJ...665L..15K} showing elongation in the moment map (bottom) and a velocity gradient in a cut along this elongation (top). The tail on NGC 4424 with \hi\ contours from WAVES (blue) and VIVA (red) overlaid on the map of \citet{2017MNRAS.464..530S}. The KW Cloud seen by \citet{2017MNRAS.464..530S}, with the WAVES data (blue) overlaid on their spectrum.}
\label{fig1}
\end{figure*}

Initial results from the first five square degrees of WAVES were published in \citet{2019AJ....158..121M}, including the discovery of around 150\% more mass in \hi\ in the ALFALFA Virgo 7 Complex \citep{2007ApJ...665L..15K} and of a much more extended ram-pressure stripped tail on NGC 4522 than was seen by \citet{2007ApJ...659L.115C}. More recent analysis of the known dark features in the rest of the WAVES area are presented in Figure \ref{fig1}. These are:
\begin{itemize}
\item{VirgoHI 4 \citep{2004MNRAS.349..922D}, where the WAVES observations reveal around 50\% more \hi\ than the WSRT observations of \citet{2005AA...437L..19O} and showing that although the \hi\ plume does extend further to the north east, where the WSRT observations reached the limits of the primary beam, this extension is only on the order of a few arcminutes (approximately the size of the Arecibo beam).}
\item{ALFALFA Virgo Cloud 3 \citep{2007ApJ...665L..15K}, which was noted as unresolved in the ALFALFA data but with the greater sensitivity of AGES is seen to be elongated (with a position angle of around $62.5\deg$) and to show a clear velocity gradient.}
\item{\hi\ tail on NGC 4424 originally seen by VIVA \citep{2007ApJ...659L.115C} and with a large extension detected by \citet{2017MNRAS.464..530S}. While they have a combined WSRT and KAT-7 dataset, both with similar column density sensitivities, this extension is only seen in their KAT-7 data (see their Figure 5 and associated discussion). WAVES reaches the same column-density sensitivity as their combined map and does not detect this extension, while follow-up single-pixel observations with the L-Band Wide receiver at Arecibo similarly failed to detect it. We therefore conclude that this was probably an artefact in the KAT-7 data and is not real.}
\item{KW Cloud detected by \citet{2017MNRAS.464..530S} near NGC 4451. This is not detected by WAVES despite being well above the sensitivity limit, nor was it seen in the ALFALFA data by \citet{2007ApJ...665L..15K} where (based on the stated mass and velocity width) it should have fallen above their $6.5\sigma$ confident detection threshold. As with the extension to NGC 4424, we were also unable to detect this source in observations with the L-Band Wide receiver at Arecibo, and thus we conclude that it is also not real. While \citet{2017MNRAS.464..530S} do not state whether this was detected in both the WSRT and KAT-7, the \hi\ morphology shown in their Figure A2 is similar to that of the extension on the \hi\ tail of NGC 4424, leading us to speculate that this is also an artefact in the KAT-7 data.}
\end{itemize}

\section{Conclusions}

WAVES is revealing new details about \hi\ in the Virgo cluster and verifying (and sometimes falsifying) earlier detections of optically-dark \hi\ clouds. We expect full analysis of the data set to reveal new dark objects and ram pressure tails, as with the AGES Virgo cluster fields. It is able to do this thanks to the combination of high brightness-temperature sensitivity, zero-spacing information, high mapping speed and relatively high spatial resolution that was offered by the Arecibo telescope. The loss of that telescope leaves the US community with limited options for such surveys that are sensitive to low column-density gas.

\begin{acknowledgments}

This research was conducted in part at the SOFIA Science Center, which is operated by the Universities Space Research Association under contract NNA17BF53C with the National Aeronautics and Space Administration. RT and BD acknowledge the support of the Czech Science Foundation grant 19-18647S and the institutional project RVO 67985815. The Arecibo Observatory is a facility of the National Science Foundation operated under cooperative agreement by the University of Central Florida and in alliance with Universidad Ana G. Mendez, and Yang Enterprises, Inc.

\end{acknowledgments}

\end{document}